\documentclass{emulateapj}

\begin{document}

\title{Extended Mid-Infrared Aromatic Feature Emission in M~82}

\shorttitle{Extended MIR Emission in M~82}

\author{C.~W.\ Engelbracht\altaffilmark{1}, P.\ Kundurthy\altaffilmark{1},
K.~D.\ Gordon\altaffilmark{1}, G.~H.\ Rieke\altaffilmark{1}, R.~C.\
Kennicutt\altaffilmark{1,2}, J.-D.~T.\ Smith\altaffilmark{1}, M.~W.\
Regan\altaffilmark{3}, D.\ Makovoz\altaffilmark{4}, M.\ Sosey\altaffilmark{3},
B.~T.\ Draine\altaffilmark{5}, G.\ Helou\altaffilmark{4}, L.\
Armus\altaffilmark{4}, D.\ Calzetti\altaffilmark{3}, M.\
Meyer\altaffilmark{3}, G.~J.\ Bendo\altaffilmark{6}, F.\
Walter\altaffilmark{7}, D.\ Hollenbach\altaffilmark{8}, J.~M.\
Cannon\altaffilmark{7}, E.~J.\ Murphy\altaffilmark{9}, D.~A.\
Dale\altaffilmark{10}, B.~A.\ Buckalew\altaffilmark{4}, and K.\
Sheth\altaffilmark{4}}

\altaffiltext{1}{Steward Observatory, University of Arizona, Tucson, AZ 85721;
cengelbracht@as.arizona.edu}
\altaffiltext{2}{Institute of Astronomy, University of Cambridge, Madingley
Road, Cambridge CB3 0HA, UK.}
\altaffiltext{3}{Space Telescope Science Institute, 3700 San Martin Drive,
Baltimore, MD 21218}
\altaffiltext{4}{Spitzer Science Center, California Institute of Technology,
Pasadena, CA 91125}
\altaffiltext{5}{Princeton University Observatory, Peyton Hall, Princeton, NJ
08544}
\altaffiltext{6}{Astrophysics Group, Blackett Laboratory, Imperial College of
Science Technology and Medicine, Prince Consort Road, London SW7 2BZ, UK.}
\altaffiltext{7}{Max-Planck-Institut f\"{u}r Astronomie, K\"{o}nigstuhl 17,
D-69117 Heidelberg, Germany}
\altaffiltext{8}{NASA Ames Research Center, Moffett Field, CA 94035}
\altaffiltext{9}{Department of Astronomy, Yale University, P.O. Box 208101,
New Haven, CT 06520-8101}
\altaffiltext{10}{Department of Physics and Astronomy, University of Wyoming,
Laramie, WY 82071}

\begin{abstract}

We present new images (groundbased optical and mid-infrared (MIR) from the
{\it Spitzer Space Telescope}) and spectra (from {\it Spitzer}) of the
archetypal starburst galaxy M~82.  The {\it Spitzer} data show that the MIR
emission extends at least 6~kpc along the minor axis of the galaxy.  We use
the optical and infrared data to demonstrate that the extended emission is
dominated by emission from dust.  The colors of the MIR emission and the
spectra indicate that there is a strong component of aromatic feature emission
(the MIR features commonly attributed to polycyclic aromatic hydrocarbons).
The dust continuum and aromatic feature emission are both strong in the
well-known superwind region of this galaxy; clearly the carrier of the
aromatic features can survive in close proximity to the wind, far from the
plane of the galaxy.  We also see significant emission by dust well outside
the superwind region, providing the clearest picture to date of the dust
distribution in the halo of this galaxy.

\end{abstract}


\section{Introduction}

M~82 is a nearby starburst galaxy in the M~81 group, and perhaps the most
well-studied member of the starburst class.  It has a high luminosity
\citep[$5\times10^{10}L_{\sun}$;][]{mcleod93}, most of which is emitted in the
infrared.  It is the brightest far-infrared (FIR) extragalactic source in the
sky after the Magellanic clouds.  It is one of the first galaxies to be
identified as a starburst galaxy \citep{rieke80}.  As an archetypal starburst
galaxy, M~82 has been observed often and over a wide range of wavelengths.  An
incomplete sampling of recent studies that can be used as a guide to previous
observations includes:  X-ray \citep{strickland04}, ultraviolet
\citep[UV;][]{hoopes05}, optical and near-infrared
\citep[NIR;][]{mccrady05,alonso-herrero03}, mid-infrared
\citep[MIR;][]{foerster-schreiber03a}, FIR \citep[][]{colbert99}, and radio
\citep{rodriguez04}.  The galaxy is also a popular target for stellar
population modeling efforts
\citep[e.g.,][]{rieke80,rieke93,satyapal97,foerster-schreiber03b}.  M~82 is
interacting with the spiral galaxy M~81 and the nearby dwarf starburst galaxy,
NGC~3077, as evidenced by the HI tails linking the three galaxies
\citep{yun94}.  M~81 lies at a distance of 3.6~Mpc \citep{freedman01}, and we
adopt this distance for M~82 as well.

M~82 is a so-called ``superwind'' galaxy in which the central starburst drives
a mass outflow perpendicular to the plane of the galaxy, as discussed in
detail by \citet{heckman90}, \citet{shopbell98}, and \citet{martin98}.  The
superwind emanates from the central starburst region, and is most evident in
X-ray \citep[e.g.,][]{strickland04} and H$\alpha$ (e.g., 
\citealp{lehnert99}) emission from the gaseous component of the wind.  There
is also dust in the superwind region, as observed in reflection \citep[][and
references therein]{hoopes05}, absorption \citep{heckman00}, and emission from
cool grains \citep{alton99}.  This work not only provides the first direct
detection of the warm dust component throughout the superwind region, but
demonstrates that the dust is distributed all around the galaxy, well beyond
the cone defined by the superwind.

Galaxies not dominated by an active nucleus in the MIR generally show a series
of strong, broad emission features arising from aromatic hydrocarbons (cf.\
\citealp{laurent00}; \citealp{lu03}; Smith et~al.\ 2006, in prep.) if
the galaxy metallicity is above $\sim1/5$ solar \citep{engelbracht05}.  These
features are often attributed to polycyclic aromatic hydrocarbons (PAHs), but
since the identification of the carrier remains controversial, we will use the
more generic term aromatic feature emission (AFE).  The features are strong in
the high-surface-brightness central regions of M~82, where they have been
observed both from the ground \citep{roche91,normand95} and from space
\citep{sturm00,foerster-schreiber03a}.  We will demonstrate that that these
features can also be found throughout the halo, up to 6~kpc from the plane of
the galaxy.

\section{Observations and Data Reduction}
\label{sec:data}

\begin{figure}
\figurenum{1}
\epsscale{0.8}
\plotone{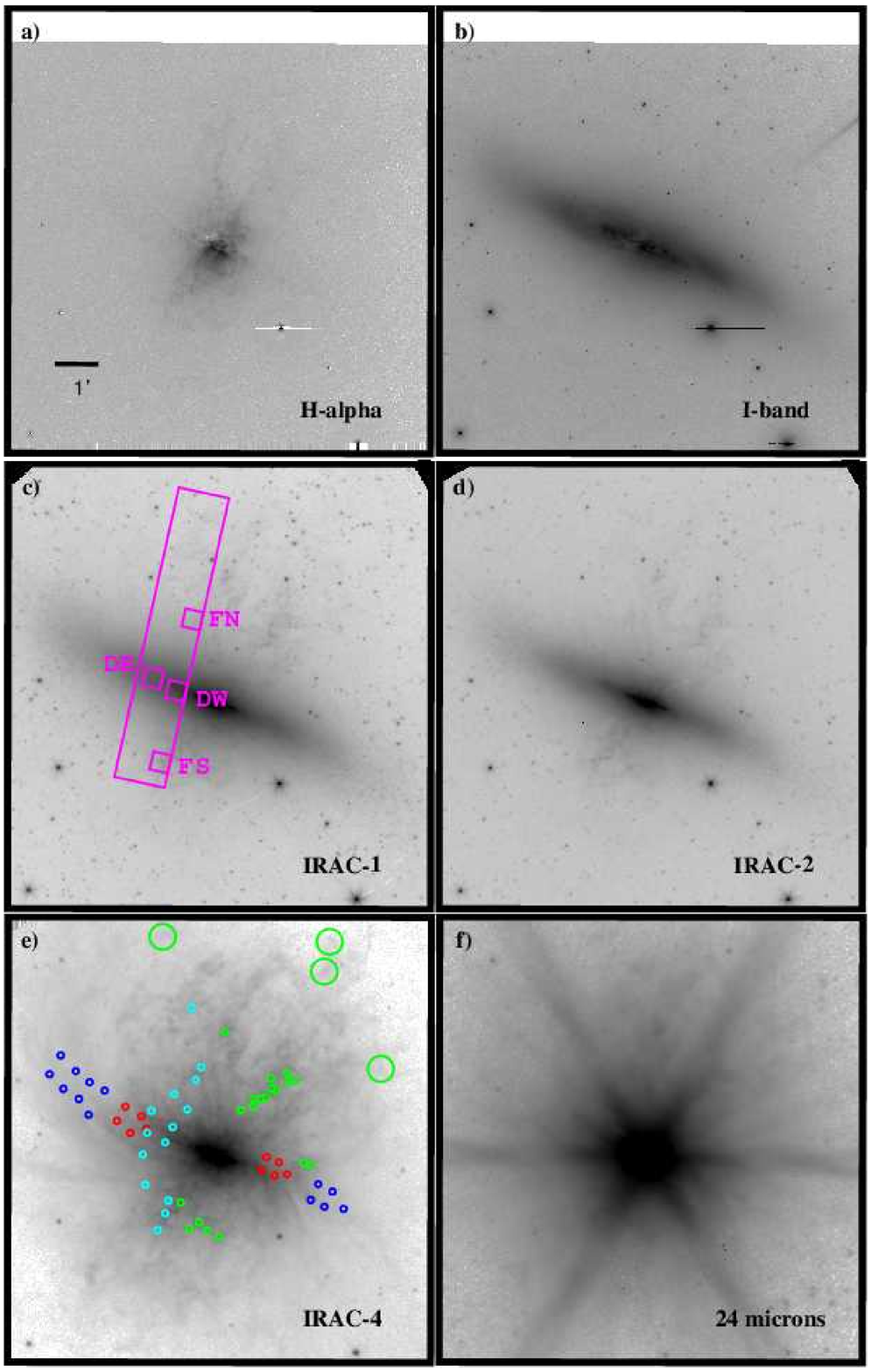}
\caption{Images of M~82, displayed in a logarithmic strech.  The images
are (a) H$\alpha$, (b) I band, (c) 3.6~\micron, (d) 4.5~\micron, (e)
8.0~\micron, and (f) 24~\micron.  Each panel is $9\farcm2\times10\farcm2$ on a
side, or $9.6\times10.7$~kpc at the adopted distance to M~82 of 3.6~Mpc.
Sizes and locations of photometry apertures are indicated in panel (e),
while the extent of the IRS spectral map and the positions of 4 spectroscopic
apertures are labeled in panel (c).}
\label{fig:images}
\end{figure}

The data presented in this paper were obtained as part of the {\it Spitzer}
Infrared Nearby Galaxy Survey \citep[SINGS;][]{kennicutt03}.  We employ here a
subset of the suite of SINGS data which allows us to trace stars (I band),
ionized gas (continuum-subtracted H$\alpha$), and hot dust ($3-9$~\micron\
images from IRAC, a 24~\micron\ image from MIPS, and $14-36$~\micron\ spectra
from IRS) in M~82.  To minimize detector artifacts due to the high surface
brightness of the central starburst, the IRAC data were obtained in separate
maps for the inner and outer regions.  Both the IRAC and MIPS data required
custom reduction procedures to produce maps free of detector artifacts.  The
images are shown in Figure~\ref{fig:images}.  Note that the 24~\micron\
image has strong diffraction spikes due to the extremely bright nucleus.

\begin{figure}
\figurenum{2}
\epsscale{1.2}
\plotone{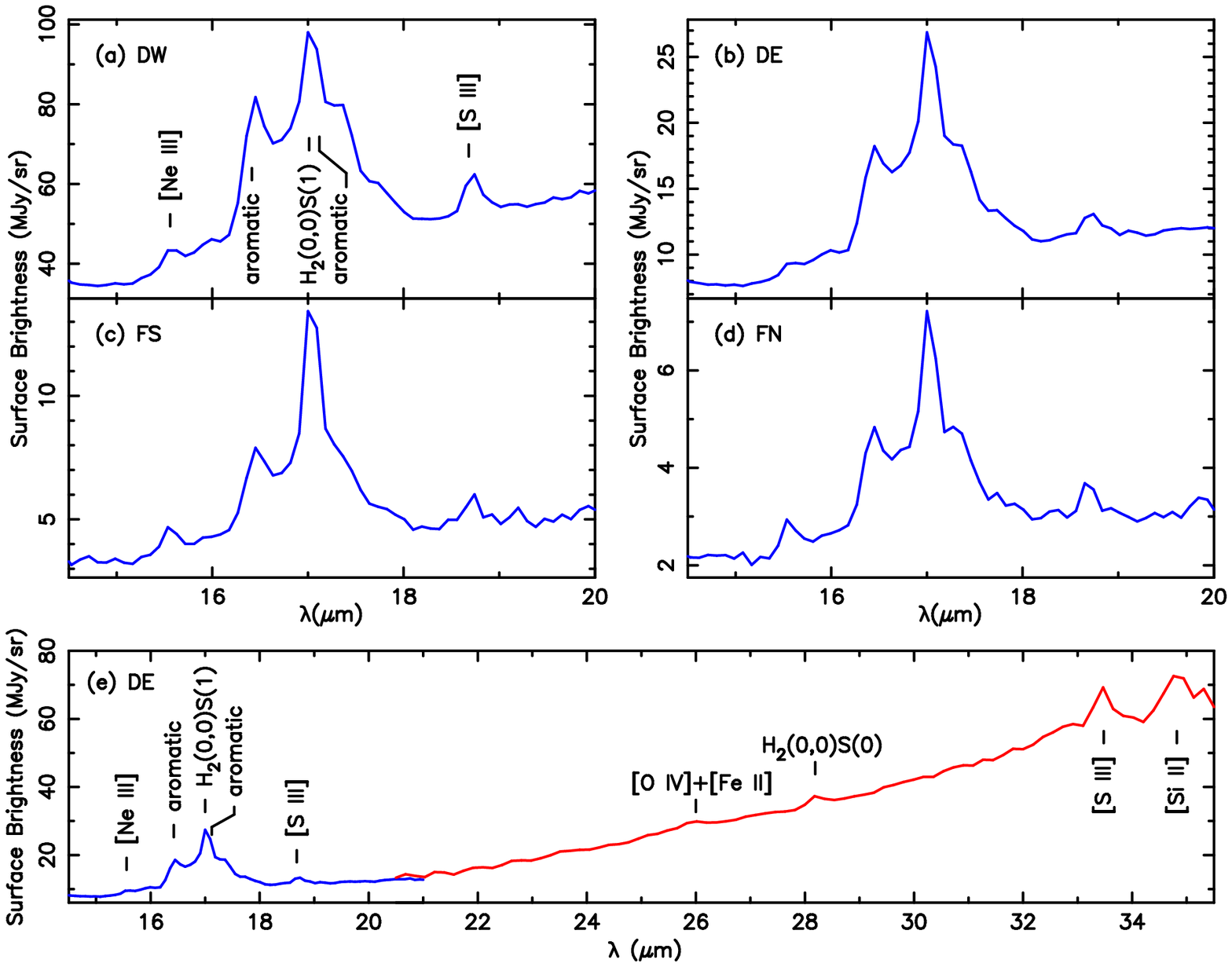}
\caption{Rest-wavelength IRS spectra of selected regions of M~82.  Panels (a)
through (d) show the portion of the spectrum around the aromatic features in
the western disk region (DW), the eastern disk region (DE), the south filament
(FS), and the north filament (FN), respectively.  Panel (e) shows the full
spectrum of the eastern disk region, where blue indicates the spectrum
extracted from long-low order 2 while red indicates the spectrum extracted
from long-low order 1.}
\label{fig:spectra}
\end{figure}

We also obtained a spectral map with IRS in a 1\arcmin-wide strip along
the minor axis of the galaxy, centered on the eastern portion of the disk.
Spectra were extracted from 4 regions (indicated in panel ``c'' of
Figure~\ref{fig:images}) where high S/N could be obtained.  These data
are shown in Figure~\ref{fig:spectra}, where we highlight the portion of
the spectrum around the aromatic features; the full spectrum of one region
is shown for reference.  The resolution ($\lambda/\Delta\lambda$) of these
data is $\sim60-130$.

We performed photometry on the images at positions indicated in panel ``e'' of Figure~\ref{fig:images}.  The regions were chosen to sample a variety
of environments, from high-surface-brightness disk regions to diffuse regions
well outside the disk of the galaxy.  Colors indicate different types of
regions: red = disk, blue = edge of disk, light blue = overlap with IRS
spectra, green = diffuse filament.  Most of the apertures are circular, with a
radius of 4\farcs25 or 18\arcsec\ (which can be compared to the HWHM of
the IRAC and MIPS beams of $\sim$1\arcsec\ and 3\arcsec, respectively) chosen
to encompass typical emission regions without overlapping.  The regions
measured contain mostly diffuse emission and so the only aperture corrections
made were to the IRAC surface brightness \citep{reach05}.  To minimize the
impact of the bright nucleus on our 24~\micron\ measurements, we first
subtracted a simple model of the nuclear region \citep[based on a 20~\micron\
map by][]{rieke80}, convolved with the 24~\micron\ PSF generated using 
observations of HD~44179 and scaled to
match the observed emission in the non-saturated regions of the galaxy.  The
subtracted image is shown in Figure~\ref{fig:colorimage}.  In all cases, a
background value measured in regions free of galaxy emission, outside the
area shown in Figure~\ref{fig:images}, was subtracted from the measurements.

\section{Results}
\label{sec:results}

\subsection{Morphology}
\label{sec:morphology}

\begin{figure}
\figurenum{3}
\plotone{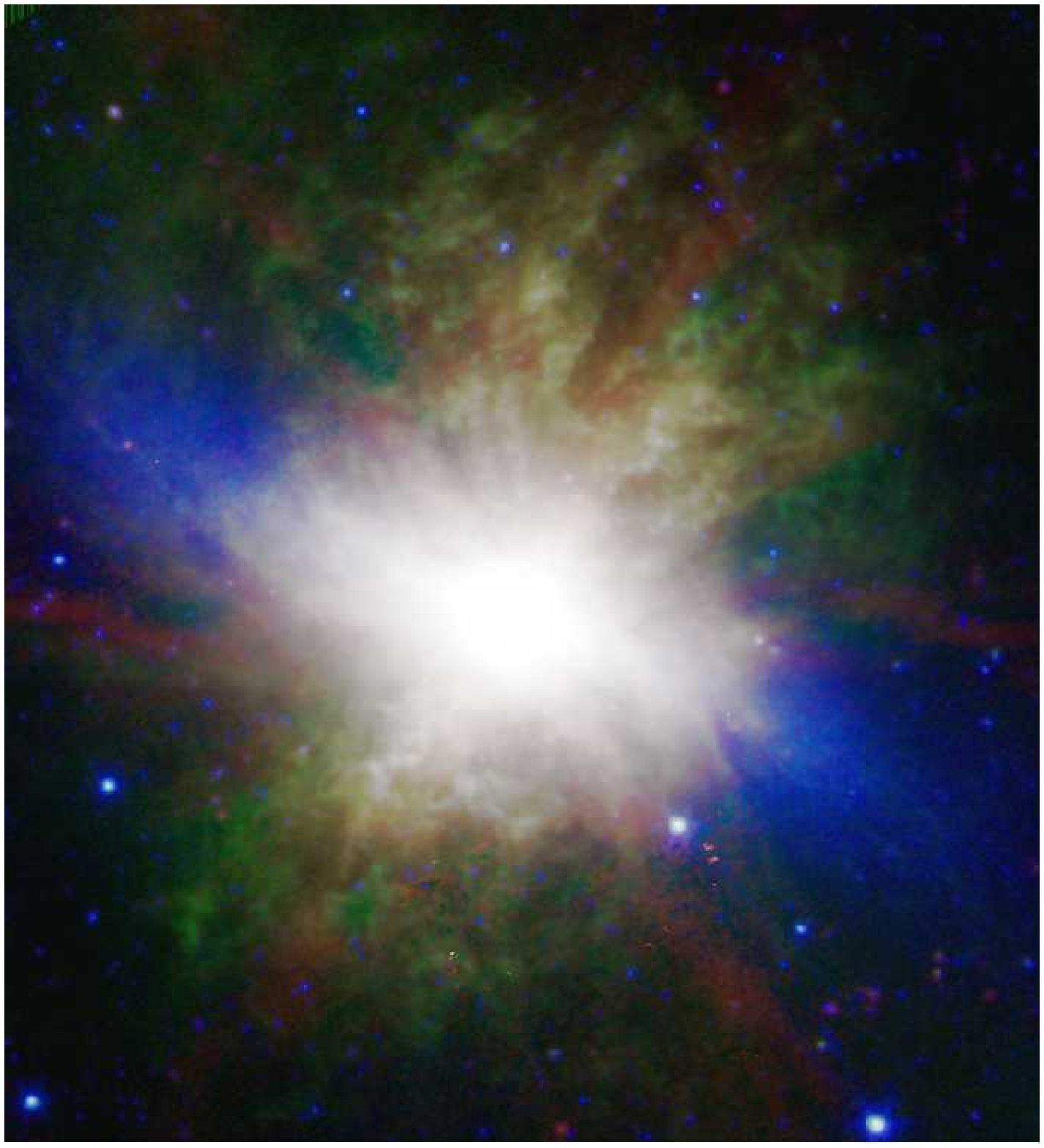}
\caption{Color composite image of M~82, displayed in a logarithmic stretch.
The blue channel is the IRAC band 1 image at 3.6~\micron, the green channel is
the IRAC band 4 image at 8.0~\micron, while the red channel is the MIPS band 1
image at 24~\micron, where the bright nucleus and associated diffraction
spikes have been subtracted as described in \S\ref{sec:data}; the radial red
streaks are residuals of the subtraction.  The angular
extent of the image is the same as the panels in Figure~\ref{fig:images}.}
\label{fig:colorimage}
\end{figure}

The morphology of M~82 is a strong function of wavelength.  The optical
broadband images are dominated by an edge-on disk with a prominent dust
absorption lane near the center.  This dust lane conceals the nucleus, near
the peak of the starburst activity.  This becomes obvious at longer
wavelengths less affected by extinction: in the IRAC band~1 image at
3.6~\micron, the prominent disk is more symmetric and dominated by a central
bright source similar to groundbased K-band images
\citep[e.g.,][]{mcleod93,jarrett03}.  It is in IRAC band~1, though, that we
begin to see a departure from the picture obtained from the optical images ---
there is filamentary emission perpendicular to the disk of the galaxy which
extends for several arcminutes (an arcminute is roughly a kiloparsec at the
distance of M~82).  This emission is more prominent at 4.5~\micron, and
becomes a dominant characteristic of the image by 8.0~\micron, where filaments
are detected a full 6~kpc from the disk of the galaxy.  The extended emission
observed in the IRAC bands accounts for a significant fraction of the total:
$\sim1/3$ of the 4.5 and 8.0~\micron\ emission arises outside the central
$1\times0.6$~kpc, while the equivalent fraction at 24~\micron\ is $\sim1/5$
(where we have used the model described in \S\ref{sec:data} to account
for the saturated region).  The diffuse extraplanar MIR emission evokes the
H$\alpha$ image, which is dominated by a biconical structure that defines the
superwind.  In the superwind region, the morphology of the MIR emission is
similar to that of the H$\alpha$ emission and other tracers of the superwind
such as UV \citep{hoopes05} and X-ray \citep{strickland04}, but the MIR images
differ in that the emission is bright all around the galaxy rather than being
dominated by a cone perpendicular to the plane.  Although the morphology at
24~\micron\ is dominated by the very bright central source and the
accompanying diffraction spikes, between the diffraction spikes the diffuse
emission is very similar to the 8~\micron\ image, as shown in
Figure~\ref{fig:colorimage}.  Clearly, the infrared images reveal a different
component of the galaxy than the optical or superwind tracers.

The extent of the MIR emission is highly unusual.  Inspection of other images
in the SINGS sample or in published {\it Spitzer} samples
\citep[e.g.,][]{pahre04} does not reveal any other galaxy with similar
characteristics.  We find MIR emission extending along the minor axis of the
galaxy from all parts of the disk, not just the superwind region.  This
suggests that there is a material in the halo of this galaxy that is capable
of significant emission in the MIR.  The material may have been ejected from
the disk over time via a galactic fountain process similar to, but more
vigorous than, NGC~5907 \citep{irwin06}, but the unique nature of the MIR
emission from the halo of M~82 argues against such a general process.  More
likely, the interaction with M~81 may have ejected material over a
shorter time by triggering a starburst (e.g., \citealp{degrijs01};
\citealp{mcleod93}).  Some of this material may be entrained in the superwind
as in NGC~253 \citep[][]{tacconi05}, but it is likely much of the MIR emission
traces a medium with which the superwind is interacting as it expands into the
halo.

\subsection{Components of Extraplanar MIR Emission}
\label{sec:components}

We turn now to elucidating the nature of the extraplanar MIR emission:
dust emission, starlight, or line emission, and their relative contribution.

We compute the stellar contribution at MIR wavelengths by scaling a
short-wavelength measurement (e.g., I band or 3.6~\micron) by the appropriate
factor, derived using stellar population synthesis models.  In the disk, where
the extinction is high and the equivalent width of the 3.3~\micron\ aromatic
feature is low \citep[cf.][]{foerster-schreiber01}, we scale the 3.6~\micron\
band by factors 0.57 and 0.23 at 4.5~\micron\ and 8.0~\micron, respectively
\citep[cf.][]{helou04,engelbracht05}, to compute the stellar contribution.  In
the halo, where the extinction is low and the equivalent width of the
3.3~\micron\ aromatic feature is possibly high (as evidenced by the
extraplanar filaments in the 3.6~\micron\ image shown in
Figure~\ref{fig:images}), we scale the I-band by factors 0.18 and 0.075 (as
computed using the model by \citealp{leitherer99}, about a factor of 2 lower
than we obtain using the model by \citealp{bruzual03}) at 4.5~\micron\ and
8.0~\micron, respectively.  We find that the stellar contribution to the
emission well outside the plane of the galaxy is no more than a few percent in
any IRAC band.  In all cases, only a negligible fraction of the 24~\micron\
emission is stellar.

The only prominent spectral lines typically found in galaxies that fall in the
MIR bands we consider here that are not due to AFE are both found in the IRAC
4.5~\micron\ band:  Br$\alpha$ at 4.05~\micron\ \citep[observed in the
nucleus of M~82;][]{foerster-schreiber01} and multiple H$_2$ lines
\citep[e.g.,][Draine 2006, in prep.]{noriega04}.  We compute the Br$\alpha$
contribution by scaling the observed H$\alpha$ emission by the hydrogen
recombination line intensities computed by \citet{storey95}.  This estimate is
subject to uncertain extinction corrections in the dusty disk of the galaxy,
but should be robust for the extraplanar emission.  We find that Br$\alpha$
contributes no more than a few percent of the total 4.5~\micron\ emission in
any of our photometric regions.  We estimate the strength of the H$_2$
lines in the 4.5~\micron\ band by scaling the strength of the (0,0)S(0) line
according to the PDR models by \citet{draine96}.  We find the possible H$_2$
contribution at 4.5~\micron\ covers a wide range, from $\sim$0.001 to nearly 30
percent.

\begin{figure}
\figurenum{4}
\epsscale{1.2}
\plotone{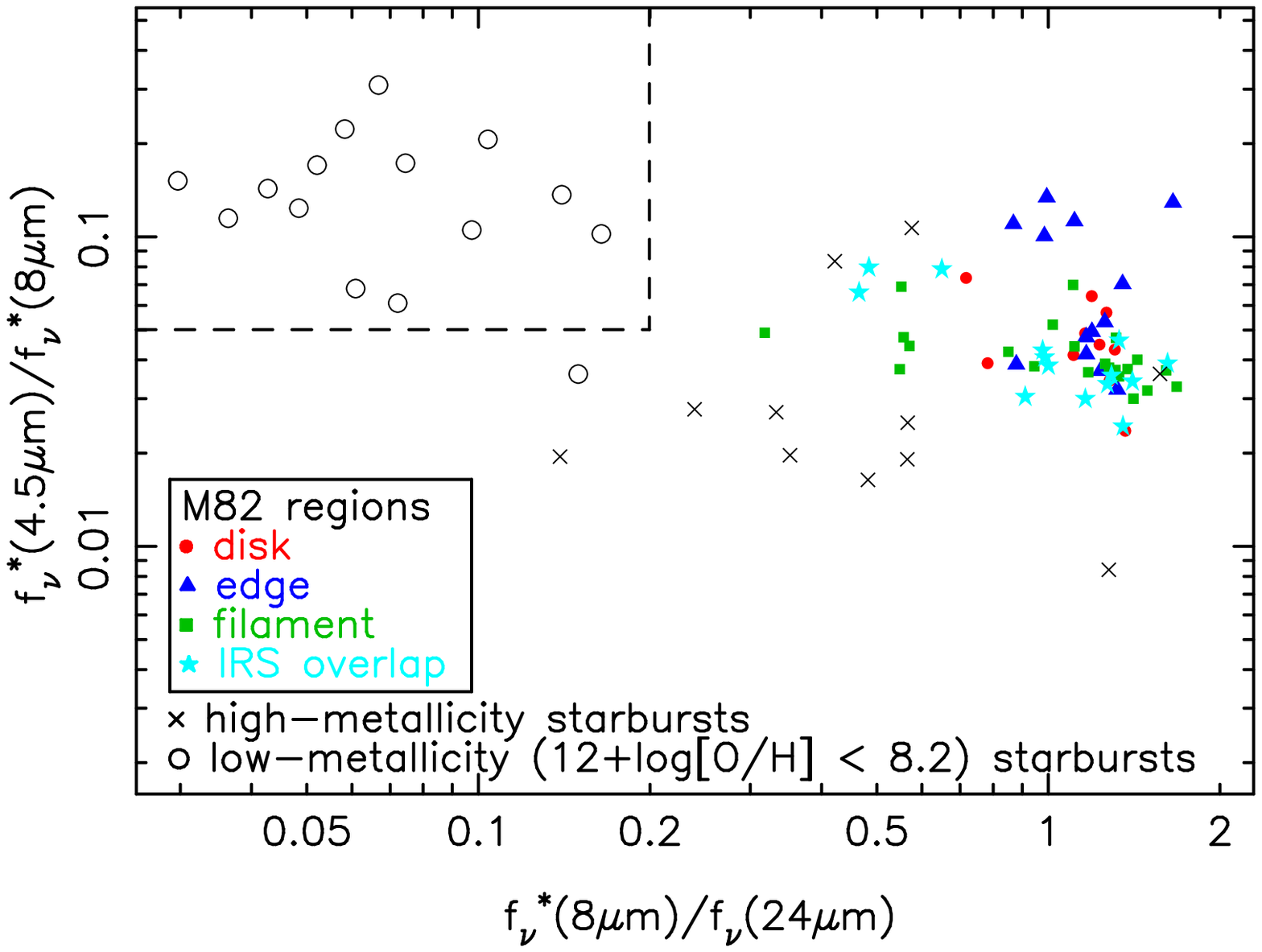}
\caption{Colors of selected regions of M~82.  A ``*'' indicates that stellar
emission has been subtracted as described in the text.  Also plotted are the
starburst galaxy points from \citet{engelbracht05}.  The dashed line
encloses the color space found by those authors to indicate a lack of AFE.}
\label{fig:color-color}
\end{figure}

Thus, the contribution to the extraplanar emission from starlight and hydrogen
emission lines is insignificant (except perhaps the H$_2$ contribution at
4.5~\micron).  We conclude that the extraplanar MIR emission is dominated by
dust.  At 3.6~\micron, the emission is likely a combination of AFE and hot
dust \citep[cf.][]{helou00,lu03}.  We turn to broadband colors and
spectroscopy to determine what kind of dust (specifically, continuum or AFE)
is contributing to the 8~\micron\ band, starting with the diagnostic diagram
developed by \citet{engelbracht05}.  Our photometric regions are plotted in
Figure~\ref{fig:color-color}, where we see that all the points fall in the
part of the plot where the 8~\micron\ band is dominated by AFE.  This is
consistent with our spectra, which show the 17~\micron\ AFE feature discovered
by \citet{smith04} and \citet{werner04} remains strong even in the extraplanar
regions (see Figure~\ref{fig:spectra}).  The strength of the 17~\micron\
complex, scaled by the average (8~\micron\ complex)/(17~\micron\ complex)
ratio of SINGS galaxies (Smith et~al.\ 2006, in prep.), is sufficient to
explain the IRAC band~4 emission in the regions where we extracted spectra.
The fact that the dust in the outflow is readily detected in the IRAC bands
and at 24~\micron\ supports the $\sim37$K ``warm'' model of \citet{alton99}
for the extraplanar dust. This model predicts a total outflowing dust mass of
$\sim10^6$~M$_\odot$, which they show is roughly compatible with other
estimates (e.g., from reddening of background galaxies).

\section{Conclusions}

We present new {\it Spitzer} images and spectra of the starburst galaxy M~82
in the MIR and new groundbased optical imaging, including a deep H$\alpha$
map.  The {\it Spitzer} data demonstrate the galaxy has copious MIR emission
well outside the plane of the galaxy.  At 8.0~\micron, this emission extends
to 6~kpc from the plane and the emission outside the central $1\times0.6$~kpc
accounts for $\sim1/3$ of the total flux.  We use the optical data to quantify
the small contributions to the MIR bands from starlight and line emission, and
conclude that most of the MIR emission must be due to dust.  At 24~\micron,
this emission is due to a featureless continuum arising from warm dust grains.
At 3.6 and 8.0~\micron, the colors and the spectra confirm a significant
contribution by aromatic features, while the 4.5~\micron\ emission may be due
to a combination of hot dust and H$_2$ emission.

The MIR emission, especially at 8.0~\micron, is filamentary and extends in all
directions and from all parts of the disk of the galaxy.  The fact that
significant emission by dust is detected outside the well-known superwind
emanating from this galaxy suggests that some process prior to the current
starburst resulted in the expulsion of dust from all parts of the disk.  The
superwind is expanding into this dusty medium, and both small dust particles
and the carrier of the AFE appear to survive the interaction.  Some dust may
be entrained in the wind, and dust is found well beyond the radius where gas
is thought to be escaping the galaxy \citep[cf.][]{shopbell98} --- if this
dust is escaping as well, the carrier of the AFE is is entering the
intergalactic medium.

There may be alternative possibilities to the MIR-emitting dust being
entrained in the superwind and sharing its high velocity. The superwind
region is complex, containing in addition to the hot plasma, shock-heated
[\ion{Fe}{2}] emission \citep{alonso-herrero03}; molecular gas
\citep{nakai87,loiseau90,walter02}, including cool H$_2$ emitting from the
low-lying transitions seen at 17 and 28~\micron\ (this work); dust grains seen
here at 24~\micron\ and in the submm \citep{alton99}; and aromatic molecules
(this work). It is conceivable that some of the dust or aromatic molecules
form {\it in situ}, or that radiation pressure aids the escape of small
particles from the galaxy.

\acknowledgments

We thank the anonymous referee for a constructive report which improved this
paper.  This work is based in part on observations made with the {\it Spitzer
Space Telescope}, which is operated by the Jet Propulsion Laboratory,
California Institute of Technology under NASA contract 1407. Support for this
work, part of the {\it Spitzer Space Telescope} Legacy Science Program,  was
provided by NASA through Contract Numbers 1224769 and 1255094 issued by
JPL/Caltech.

\clearpage

\clearpage

\end{document}